\documentstyle[12pt,epsf]{article}
\title{ 
Survival Probability of a Mobile Particle in a Fluctuating Field}
\author{Satya N. Majumdar$^1$ and Stephen J. Cornell$^2$\\
{\small 1. Tata Institute of Fundamental Research, 
Homi Bhabha Road, Mumbai 400 005, India}\\
{\small 2. Department of Theoretical Physics, The University, Manchester,
M13 9PL, United Kingdom}
}

\begin{document}
\maketitle
\begin{abstract}
We study the the survival 
probability $P(t)$ upto time $t$, of a test
particle moving in a fluctuating external field. The particle moves according
to some prescribed deterministic or stochastic rules and survives as long as 
the external field that it sees at its own location does not change sign.
This is a natural generalization of the ``static persistence" (when
the particle is at rest) that has generated considerable recent interests.
Two types of motions of the particle are considered. In one case,
the particle adopts a strategy to live longer and in the other it just
diffuses randomly. Three different external fields were considered: (i)
the solution of diffusion equation, (ii) the ``colour" profile of the 
$q$-state Potts model undergoing zero temperature coarsening dynamics and 
(iii) spatially uncorrelated Brownian signals. In most cases studied, 
$P(t)\sim t^{-\theta_m}$ for large $t$. The exponent $\theta_m$ is
calculated via numerically, analytically by approximate methods and
in some cases exactly. It is shown in some special cases that the 
survival probability of the mobile particle is related to the persistence
of special ``patterns" present in the initial configuration of a phase
ordering system.
\end{abstract}

\vspace{0.5cm}


\newpage

\section{Introduction}

Considerable interests have been generated recently in understanding the
statistics of first passage events in spatially extended nonequilibrium 
systems. These systems include the Ising or Potts model undergoing zero 
temperature phase ordering dynamics\cite{derrida1,derrida2,derrida3}, 
simple diffusion 
equation with random initial conditions\cite{diffusion,watson}, several 
reaction-diffusion
systems\cite{krapiv1,cardy,monthus} and fluctuating interfaces either in
the steady states or approaching steady
states starting from random initial configurations\cite{krug}. Typically 
one is interested in persistence, i.e., the
probability $P_0(t)$ that at a fixed point in space, the quantity
$sign[\phi(x,t)-\langle \phi(x,t)\rangle]$ (where $\phi(x,t)$ is a 
fluctuating field
e.g., the spin field in the Ising model or the height of a fluctuating
interface) does not
change upto time $t$. In all the examples mentioned above, this 
probability decays as a 
power law, $P_0(t)\sim t^{-\theta_0}$, where the exponent $\theta_0$ is
nontrivial. This nontriviality is due to the fact that the
effective stochastic process in time at a fixed point in space becomes
non-Markovian due to the coupling to the neighbours. For a 
non-Markovian process, calculation of any 
history dependent quantity such as persistence is extremely hard barring a 
few special cases\cite{slepian, reviews}. The exponent $\theta_0$ has
also been measured in a recent experiment on a liquid crystal system
which has the same dynamics as the $T=0$ Ising model in $2$-d\cite{yurke}.
The experimental value was in good agreement with the analytical prediction
of $\theta_0$ in $2$-d Ising model\cite{majsire}. The exponent $\theta_0$
has also been measured in a recent experiment on two dimensional soap 
froth\cite{Tam}.

In the above, one studied the persistence of a {\em single} spin (e.g., in
the Ising or Potts model) of the initial random configuration. A natural
generalization of this would be to study the persistence of a {\em pattern},
and not just a single spin, present in the initial configuration. Persistent
patterns are quite abundant in nature. Examples include persistent eddies
and vortices in turbulence, the great red spot of Jupiter or certain
patterns of stock prices in financial markets. Another example is the so 
called activity-centered pattern in a self-organized system such as
an interface in a random medium\cite{Supriya} and also in certain
models of evolution\cite{Bak}. A natural question 
then is: what is the probability that a given pattern survives upto time 
$t$? 

Such persistent patterns exist also in phase ordering systems such as
the $q$-state Potts model. For example, one such pattern is an original
domain of a specific colour present in the random initial configuration
of the Potts model. One can then ask: what is the survival probability 
of such a domain upto time $t$? This quantity for the $1$-d
Potts model has recently been studied by Krapivsky and 
Ben-Naim\cite{krapiv2}. However this can be a more general question for 
any fluctuating field such as the solution of diffusion equation with
random initial configuration or a fluctuating interface approaching the 
steady state. In such examples, a domain would be a connected set of 
points where the sign of the fluctuating field is positive (or 
negative). Another example of ``pattern" persistence would 
be to study the probability that two adjacent domains in the initial 
configuration both survive upto time $t$. In this paper we develop
a general framework to study the persistence of patterns of a 
fluctuting field and discuss a few examples in detail where explicit
results can be obtained.

The general framework to study some of these ``pattern" persistence problems
consists of monitoring the motion of an external test particle launched
in the fluctuating field. The dynamics of the test particle is suitably 
chosen so that the particle evaluates where the specific pattern of the
fluctuating field is 
and moves there. The persistence of the pattern is then precisely the
survival probability of the test particle. This led us naturally to study 
a more general ``persistence of a mobile particle (P) in a field (F)"
problem (henceforth the PF problem), special cases of which
correspond to the ``pattern" persistence in the underlying field.
In this paper we study in detail a few 
examples of this general PF problem and find very rich, though often 
nonuniversal, behaviour. 

The general PF problem can be defined as follows. 
Let us consider a field $\phi(x,t)$ that fluctuates 
in both space and time. 
For example, the field $\phi(x,t)$ may be the solution of the simple 
diffusion equation ${\partial}_t\phi = \nabla^2\phi$, or the order parameter
profile of the Ising model undergoing $T=0$ coarsening
dynamics or the height profile of a fluctuating interface or may even be 
spatially uncorrelated Brownian 
signal $\partial_t\phi=\eta$ where $\eta(x,t)$ is spatially and temporally
uncorrelated Gaussian white noise. A test particle is launched at an
arbitrary initial point at 
time $t=0$. The particle moves according to some prescribed 
deterministic or stochastic rules which in general depend on the
local field profile. We now ask the question: given
the dynamics of the particle, what is the probability $P(t)$ that
the field seen by the particle at its own location does not change
sign upto time $t$?

The survival probability of a mobile particle in a field has been
studied before in the context
of heterogeneous reaction-diffusion systems\cite{krapiv1,monthus} where the
test particle was an external impurity diffusing through a 
homogeneous background. These studies dealt with a special case of our 
general PF problem, namely when the test particle is
a simple random walker and the field is the coarsening ``colour" 
field of the $q$-state Potts model at $T=0$\cite{krapiv1,monthus}.
Amongst other studies on the similar line was the computation of the 
trapping time distribution of a diffusing tracer 
particle on a solid-on-solid surface\cite{jk}. In this paper, we extend 
these studies to several other examples 
arising naturally in the context of our general PF problem.

In the PF problem, one needs to specify the dynamics 
of the field as well 
as that of the test particle. For the field, we will
consider three different cases, namely when the fields are:  (i) the solution
of diffusion equation, (ii) the spin profile of the Ising model or in general
the ``colour" profile of the $q$-state Potts model undergoing $T=0$ 
coarsening dynamics and (iii) spatially uncorrelated Brownian signals. The
motion of the test particle, in general, consists of two separate moves. 
In the first part of the motion the particle sees the local field profile
and then adopts a strategy to move in such a way so
that it can live longer. This is the ``adaptive" part of the motion
that depends on the local 
field profile. In addition to this ``adaptive" move the particle
in general may be subjected to an external noise which consitutes the
second part of the motion. This ``noisy" part consists of Brownian
moves of the particle that is independent of the local field profile. 
While the motion consisting of both ``adaptive" and ``noisy" moves
is more general, for the sake of simplicity we will restrict ourselves
to two separate cases when the motion is either purely ``adaptive" or
purely ``noisy". In 
the first case, the survival probability is
larger than the ``static" case (when the particle is at rest) and in the 
second case, it is  
smaller. In cases where survival probabilities both in the mobile and 
the static cases decay as power laws, 
the exponent inequality $\theta_{ad} \leq \theta_0 \leq \theta_{rw}$ holds
where $\theta_{ad}$ and $\theta_{rw}$ are the persistence exponents 
associated with the ``adaptive" and the ``random walk" motion of the
test particle. The relationship between the survival probability of
the test particle and the persistence of ``patterns" in the 
underlying field is established wherever possible.
In all these studies, we will restrict ourselves to
one dimension although in most cases the generalizations to higher
dimensions is quite straightforward.

The paper is organized as follows. In section-1, we consider the
``adaptive" motion of the test particle. Three cases of the fields are 
considered. By suitably choosing the adaptive strategy, 
the survival probability of the particle is related to that of a 
``pattern" (original domain in the Potts model or diffusion equation). An 
exact solution is presented for the special case when 
the motion of the ``adaptive" particle is ``directed" and the field is
spatially uncorrelated Brownian signals. In section-2, we consider the
``noisy" motion of the test particle where it performs simple
``random walk". Approximate 
analytical methods are developed to calculate the exponent 
characeterizing the power law decay of the survival probability of the
particle. In section-3, a special case of the survival of the
the diffusive test particle in the $1$-d Potts model is shown to be related
to the persistence of an initial pattern, namely two adjacent original
domains. This is also related to the fraction of {\em 
uncollided} domain walls at time $t$ in the $1$-d Potts model which we
study both by numerical and analytical methods. Finally, we 
conclude with a summary, some suggestions for future directions and possible 
experimental realizations or tests of our results. Some details about
a variational calculation are presented in the Appendix. 

\section{Adaptive Motion of the Test Particle}
\label{adapt}

In this section we consider the adaptive motion of the test particle in
which the particle adopts a strategy to move in a way such that it can live 
longer. The model and the strategy
is as follows. Consider a lattice with periodic boundary conditions for 
convenience. The field $\phi(i,t)$ evolves with time according to some
prescribed dynamics. A test particle is launched at $t=0$, at an
arbitrary site say the origin. Let us assume that at $t=0$, the sign of 
the field $\phi$
at the origin is positive (or negative). As time changes, the field 
$\phi(i,t)$ changes. As long as the sign of the field at the origin is 
positive (or negative), the particle does not move. When the sign changes
at the origin, the particle looks for a nearest neighbour where the field
is positive (or negative). If it finds such a neighbour it goes there.
In case there are more than one neighbours with positive (or negative)
fields, it chooses one of them at random. Then it waits there till the
sign of $\phi$ at the new site changes and then again it moves to one of 
its current neighbours and so on. If at some stage, the sign changes at 
the particle's current site
and it fails to find a neighbour with positive (or negative) field, then 
it dies. Then we ask: What is the probability $P_{ad}(t)$ that the
particle survives upto time $t$. Note that if the particle did not move
at all and stayed put at one site only, then the survival probability
$P_0(t)$ is the usual ``static persistence".

We first consider the case when the field $\phi (x,t)$ is the solution
of the simple diffusion equation $\partial_t\phi=\nabla^2\phi$, starting
from a random initial configuration of $\phi$. For this simple field, 
analytical computation of even the static persistence $P_0(t)$ turned out 
to be quite nontrivial\cite{diffusion}. In $1$-d, it was found that
$P_0(t)\sim t^{-\theta_0}$ for large $t$ where $\theta_0=0.1207 \pm 
.0005$\cite{diffusion}. The 
reason for the nontriviality once again can be traced to the fact that the 
effective Gaussian process that a static particle sees in time is
non-Markovian. Nevertheless, an ``independent interval approximation" 
(IIA) was developed
in\cite{diffusion}, which produced analytical predictions for $\theta_0$ 
for all dimensions that were extremely accurate. 

In case of the moving particle with the adaptive strategy, we carried 
out a numerical simulation. The results are presented in Fig. [1]. We
find, $P_{ad}(t)\sim t^{-\theta_{ad}}$ for large $t$, where 
$\theta_{ad}=0.091 \pm .002$, compared to $\theta_0=0.12\pm .001$. 
Thus the strategy adapted by the particle is  ``successful" in the 
sense that the exponent, and not just the amplitude, characterizing the 
decay of persistence decreases. Thus in the renormalization group 
language, the adaptive strategy is a relevant perturbation.

It is clear from above that the strategy that the
particle adopts is basically to move towards the local maximum (or minimum)
of the underlying field if the initial sign of the field that the 
particle sees is positive (or negative). By symmetry of the initial 
condition the
survival probability of the particle does not depend on the initial
sign of the field that it sees. Hence without any loss of generality it is
sufficient to consider the case when the particle moves only towards
local maximum. This observation may be 
used to develop a possible continuum approach to this problem. Let $R(t)$
denote the position of the particle measured from a fixed point in space 
and $x$ denote the co-ordinate of an arbitrary point in space measured from
the location of the particle. Then, the effective field $\psi(x,t)$ as seen
by the particle is given by, $\psi(x,t)=\phi(x+R(t),t)$. Note that $x=0$
denotes the position of the particle. Given that $\phi$ satisfies diffusion
equation, the equation of motion of $\psi(x,t)$ is given by, 
$\partial_t\psi={\nabla}^2\psi +{\dot R}(t){\partial_x 
\psi}$. We now model the adaptive strategy (namely that the particle
tries to move towards the local maximum) by
assuming that the velocity of the particle is proportional to 
the local slope of the field that the particle sees, i.e., ${\dot 
R}={\lambda}{\psi}'(0,t)$, where ${\psi}'$ denotes the derivative with 
respect to $x$ and $\lambda$ is a constant. The implication of this 
assumption is clear. If
the local slope is positive, the particle moves to the right and if the
local slope is negative, the particle moves to the left. Thus the particle 
always tries to move towards the local maximum. So, the equation 
satisfied by $\psi(x,t)$ is
\begin{equation}
\label{KPZ}
{\partial_t \psi}={\nabla}^2\psi +\lambda {\psi}'(x,t){\psi}'(0,t)
\end{equation}
which is a non-local and nonlinear KPZ type of equation. Then the adaptive
persistence in this formulation is the probability
that the local field $\psi(0,t)$ does not change sign upto time $t$.
While a continuum formulation does not make it easier to compute the
adaptive persistence exponent, but it relates the problem to the more
familiar problem of persistence of fluctuating interfaces\cite{krug}.
However we will not study this equation any further in the present paper
and will defer its discussion for future\cite{majdhar}.

We now turn to the case when the field is the ``colour" field of the
$q$-state Potts model undergoing $T=0$ temperature dynamics. At each
site of a lattice, the field can take $q$ possible colours. One starts
from a random initial configuration of colours. A site is chosen at
random and its colour is changed to one of its neighbours. This is how
the colour field evolves. A test particle is launched as usual and it
waits at its initial site till the colour of that site changes. Then it 
tries to 
find a neighbour with the same colour and if succeeds it goes to that 
neighbouring site. When it does not find any neighbour of its own colour,
the particle dies. Then the question is as before, what is the probability
$P_{ad}(t)$ that the particle survives upto time $t$. The corresponding
``static persistence" exponent has been calculated exactly for all $q$
recently\cite{derrida2}. 

Our job for calculating the adaptive persistence for the $q$-state Potts
model is simplified by making the observation that the test particle
survives as long as the original domain of the Potts model that contained
the test particle at $t=0$ survives. Thus the adaptive persistence
is precisely the
survival probability of an ``original domain" that has recently been 
studied both numerically and analytically within an independent interval 
approximation (IIA) for all $q$ by Krapivsky and Ben-Naim\cite{krapiv2}.
In fact, even for the diffusion equation, adaptive
persistence is also the survival probability of an original domain. However,
the IIA approximation developed in\cite{krapiv2} for the Potts model can 
not be easily extended to the diffusion equation for the following reason.
The evolution of the Potts model is particularly simple in terms of the 
domain walls where the field changes colour in space. These domain walls 
perform independent random walks (the rates of which do not depend on
the local spins) and when two walls meet, they either
annihilate (with probability $1/(q-1)$) or aggregate (with probability 
$(q-1)/(q-2)$). 
Therefore, for the Potts model it is quite simple to write down
an evolution equation for $P(n,m,t)$ (probability that a domain of
length $n$ contains $m$ original domains) within IIA\cite{krapiv2} and 
thereby calculate the domain survival probability. However, writing down a 
similar evolution equation for the diffusion equation does not seem to be 
easy as the domain walls in the diffusion equation (i.e., the zeroes of
the diffusive field) undergo complicated motion that depends upon the local
field profile.

However we want to stress that the concept of ``adaptive 
persistence" is more general than just being equivalent to the 
persistence of a pattern, e.g.,the domain survival probability in case 
of Potts model or diffusion equation. In 
fact, it does not necessarily require that the evolving field has domain
structures coarsening in time. For example, even in the simplest case 
where at each site of the lattice there is an independent Brownian signal 
(completely uncorrelated spatially), one can define the ``adaptive 
persistence" and has a nontrivial exponent as we will show below via an 
exact solution. In cases when the 
evolving field has a structure of coarsening domains, the adaptive 
persistence is equivalent to the domain survival probability.

We now turn to the last case for which we present an exact solution of 
the adaptive persistence. In this case, the field is spatially 
uncorrelated Brownian signal at each lattice site, $\partial_t \phi 
(i,t)=\eta(i,t)$ where $\eta(i,t)$'s are spatially uncorrelated Gaussian 
white noises with zero mean and $\langle 
\eta(i,t)\eta(j,t')\rangle={\delta}_{i,j}\delta(t-t')$. A test particle 
is launched as usual at $t=0$ at the origin. Let us assume that the
signal at the origin at $t=0$ is positive. The particle does not move as 
long as the sign of the signal $\phi$ at the origin does not change sign.
When it does, the particle either dies with probability $(1-p)$ or survives
with probability $p$ and then tries to jump to its  neighbour on the {\em 
right} hand side.
If the sign of the signal at that neighbour is positive at the time of
jumping, the particle stays there till the signal is positive there and
so on. Finally, if it does not find a {\em right} neighbour with a positive
signal at the moment of jumping, the particle dies. Note the two new
aspects in this problem from before. A survival factor $p$ is introduced.
$p=1$ is the fully adapted case considered earlier for the diffusion
equation or the Potts model. $p=0$ will correspond to the ``static
persistence". The second aspect is that the motion of the particle is
{\em directed} as opposed to the {\em undirected} case considered for
the diffusion equation or the Potts model. This assumption of
{\em directedness} turns out to be important for exact solution and 
thus serves as an useful exactly solvable toy model of adaptive
persistence. It turns out, as we show below, that the exponent
$\theta_{ad}$ can be calculated exactly and depends on the parameter $p$ 
continuously. 

Let $P_0(t,t_0)$ denote the probability that a Brownian signal at a
given site does not change sign from time $t_0$ to time $t$. This is
the usual static persistence, which can be computed exactly since
it is a Markovian process and is given by, $P_0(t_0,t)={2\over 
{\pi}}\sin^{-1}( {{min(t_0,t)}\over {\sqrt {t_0t}}})$\cite{slepian}. Let 
$F_0(t_0,t)$ denote the probability
that the signal crosses zero for the first time at time $t$. Then clearly,
$F_0(t_0,t)=-dP_0/dt$. Then the probability of no zero crossing $P(t_0,t)$
for the adaptive particle is given by the convolution,
\begin{equation}
\label{conv}
P(t_0,t)=P_0(t_0,t) +(p/2)F_0*P_0 +(p/2)^2 F_0*F_0*P_0 +\ldots,
\end{equation}
where $F_0*P_0=\int F_0(t_0,t_1)dt_1P_0(t_1,t)$ and so on. The first term
is the probability that the sign of the signal at the starting site did
not change upto time $t$ and hence the particle did not jump at all. The 
second 
term denotes the probability that the particle jumped once. The parameter
$p$ is the survival factor and $1/2$ is the probability that the 
sign of the signal of the right neighbour (to which the particle jumps)
is positive. The third term denotes the probability that the particle 
jumped twice and so on. 

To perform the convoluted integrals in Eq. (\ref{conv}) , we make a 
change of variable, $T_i=log(t_i/t_0)$.
In this new variable, the effective process that the particle sees, though
not Gaussian, becomes stationary. One can then use the Laplace transforms
to solve Eq. (\ref{conv}). Let ${\tilde P}(s)$, ${\tilde P_0}(s)$ and
${\tilde F_0}(s)$ denote the respective Laplace transforms of $P(T)$, 
$P_0(T)$ and $F_0(T)$. 
Then, by taking the Laplace transform of Eq. (\ref{conv}), one gets,
\begin{equation}
\label{laplace}
{\tilde P}(s)={{{\tilde P_0}(s)}\over {1-{p\over {2}}{\tilde F_0}(s)}}.
\end{equation}
Using the realtion, $F_0(T)=-dP_0/dT$, one further gets,
${\tilde F_0}(s)=1-s{\tilde P_0}(s)$. This enables us to write ${\tilde 
P}(s)$ entirely in terms of ${\tilde P_0}(s)$,
\begin{equation}
\label{lap}
{\tilde P}(s)={ {{\tilde P_0}(s)}\over {1-{p\over {2}}+{ps\over 
{2}}{\tilde P_0}(s)} }.
\end{equation}

We expect $P(t_0,t)$ to decay as $t^{-\theta_{ad}}$ for large $t$. This
means that in the variable $T=log(t/t_0)$, $P(T)\sim \exp
(-\theta_{ad}T)$ for large $T$. This implies that the Laplace transform
${\tilde P}(s)$ will have a pole at $s=-\theta_{ad}$, i.e., the denominator 
of the right hand side of Eq. (\ref{lap}) will have a zero at 
$s=-\theta_{ad}$. Using $P_0(T)={2\over 
{\pi}}\sin^{-1}(e^{-T/2})$, one therefore sees that that the exponent
$\theta_{ad}$ is given by the positive root of
\begin{equation}
\label{zero}
1-{p\over {2}}-{p\theta_{ad}\over 
{\pi}}\int_0^{\infty}\sin^{-1}(e^{-T/2})e^{\theta_{ad}T}dT =0.
\end{equation}
This integration can be evaluated by parts and one finally gets,
\begin{equation}
\label{theta}
B\bigl[1/2, 1/2-\theta_{ad}\bigr]= {{2\pi}\over {p}}
\end{equation}
where $B[m,n]$ is the usual Beta function. It is clear that in the
limit, $p\to 0$, one recovers the usual ``static" persistence exponent
$\theta_0=1/2$. For the fully adapted model ($p=1$), we get
$\theta_{ad}=0.3005681\dots$ which agrees very well with our numerical
simulations. It is also clear that for any nonzero $p$, the adaptive
exponent $\theta_{ad}< \theta_0=1/2$ as expected.

The reason that the exponent $\theta_{ad}$ is exactly soluble for the
{\em directed} case is that the effective process seen by the test
particle is Markovian. In the {\em undirected} case, this is not so
because the particle can jump back to a site already visited before
and therefore the probability that the signal is positive there at
the time of current jumping is conditioned by the fact that the signal
had crossed zero there at some earlier time. Therefore, it is difficult
to compute $\theta_{ad}$ exactly for the {\em undirected} case.
 
\section{Noisy Motion of the Test Particle}
\label{random}
In the previous section we considered the adaptive motion of the test
external noise. In this section we consider the other case when
there is only noise and no adaption. In this case the particle
just moves randomly through
the medium in which a field $\phi(x,t)$ is evolving according to its
own prescribed dynamics. As before, the test particle is launched
at the origin at $t=0$ where the sign of the field is positive (say)
at $t=0$. The particle then performs a Brownian motion, $\dot R =\eta (t)$,
where $\eta(t)$ is a Gaussian white noise with zero mean and the
correlator, $\langle \eta(t)\eta(t')\rangle=2D_p\delta(t-t')$. Here 
$R(t)$ denotes the position of the particle from some fixed reference 
point. The
motion of the field and that of the particle are completely uncorrelated.
The particle dies when the field that it sees changes sign. Then we
ask: what is the probability that the particle survives upto time $t$?

We first consider the case when the field $\phi(x,t)$ is evolving 
according to diffusion equation, $\partial_t\phi=D_f\nabla^2\phi$.
We performed numerical simulation to compute the survival probability
of the test particle. This probability decays as a power law, $P(t)\sim 
t^{-\theta_d}$ for large $t$ where the exponent $\theta_d$ is found to
depend continuously on the ratio of the two diffusion constants, $c=D_p/D_f$.
The results are presented in the second column of Table (1) under the
heading $\theta_d(MC)$. It is 
clear
from this table that for any nonzero $c$, $\theta_d (c)> \theta_0$ where
$\theta_0$ is the corresponding ``static" persistence exponent, i.e., when
$c=0$. This continuous non-universal dependence of $\theta_d$ on $c$ is,
however, not very surprising for the following reason. The test particle
dies whenever it crosses any ``zero" of the field $\phi (x,t)$. Thus two
zeros of the field on either side of the test particle act like two boundary 
walls. But these walls are not static. They themselves are moving as the
field $\phi (x,t)$ is evolving in time. In fact, since the typical distance
between zeroes of the diffusing field increases as $\sqrt t$, the walls
bounding the test particle are therefore diffusing as $\sqrt t$. This 
particular case is known to be marginal\cite{krapiv3} in the sense that
the exponent, characterizing the power law decay of survival probibility
of the particle, depends continuously on the ratio of the diffusion
constants of the particle and the walls.  

While this explains qualitatively why $\theta_d$ depends continuously on $c$,
it does not give any quantitative estimate of the exponent. To make progress
in that direction, we proceed as follows. Let $x$
be the coordinate of an arbitrary point in space measured from the rest
frame of the particle. Then
the field $\psi(x,t)=\phi(x+R(t),t)$ as seen by the particle evolves 
as,
\begin{equation}
\label{effective}
{{\partial \psi}\over {\partial t}}=D_f {\nabla}^2\psi +{{\partial \psi}
\over {\partial x}}\eta(t).
\end{equation}
The field $\psi$ and the noise $\eta$ are completely
uncorrelated. Note that for a given realization of the noise process
$\{\eta(t)\}$, the process $\psi(x,t)$ at a fixed $x$ as a function of $t$
is a Gaussian process. However, when the distribution of $\eta(t)$ is
also taken into consideration, $\psi(x,t)$ at a fixed $x$ no longer has a 
Gaussian distribution due to the multiplicative nature of the noise in Eq. 
(\ref{effective}). It is nevertheless useful to calculate the
the two time correlator, $C(t',t)=\langle \psi(0,t')\psi(0,t)\rangle$ that
characterizes the temporal process at the location of the particle, i.e., 
at $x=0$. This can be easily performed in the $k$ space where the solution
is given by,
\begin{equation}
\label{Fourier}
\psi(k,t)=\psi(k,0)e^{-D_fk^2t}e^{ik\int_0^t \eta(t')dt'}.
\end{equation}
We then compute $C(t',t)=\int dk \langle \psi(-k,t')\psi(k,t)\rangle$
where the $\langle \rangle$ is done over both the initial conditions of 
$\psi$ and the history of the noise $\eta$. The initial consition is 
taken to random, so that $\langle \psi(k,0)\psi(-k,0)\rangle =\Delta$,
where $\Delta$ is a constant.
Since the noise $\eta$ is Gaussian white noise, we use the property,
$\langle \exp (ik\int_{t'}^{t}\eta(t_1)dt_1\rangle =\exp [-D_pk^2|t-t'|]$.
It is then easy to see that in $d=1$,
\begin{equation}
\label{correlator}
C(t',t) = {1\over {2\sqrt \pi}}{1\over {\sqrt {D_f(t+t')+D_p|t-t'|}}}.
\end{equation}
The normalized autocorrelator, $f(t',t)=C(t',t)/{\sqrt {C(t',t')C(t,t)}}$
when expressed in terms of the variable $T=log(t)$, becomes stationary,
i.e., only a function of the time difference $|T-T'|$. Denoting, for 
convenience, this time difference by $T$, one finds that the stationary
autocorrelator $f(T)$ is given by,
\begin{equation}
\label{auto}
f(T)= {1\over { \sqrt {\cosh (T/2)+c \sinh (|T|/2)}}},
\end{equation}
where $c=D_p/D_f$ is the ratio of the two diffusion constants. 

Note that for $c=0$, $f(T)$ reduces to the ``static" autocorrelator
$f_0(T)= [\cosh (T/2)]^{-1/2}$\cite{diffusion}. However, there is an 
important
difference between $c=0$ and $c\neq 0$ cases. For $c=0$, the stochastic
stationary process, whose correlator is given by $f_0(T)$, is Gaussian.
However, for $c\neq 0$, while the process is still stationary in the
variable $T$, it is non-Gaussian as mentioned earlier. This is evident from 
Eq.(\ref{Fourier}),
since $\psi(k,t)$ is a product of two random variables, $\psi(k,0)$ and
$\exp [ik\int \eta(t')dt']$ and hence is not clearly Gaussian, even though
both $\psi(k,0)$ and $\eta (t)$ are individually Gaussian. Therefore, the
approximate method developed\cite{diffusion} for the case $c=0$, to compute 
the 
asymptotic distribution of the interval between successive zero crossings,
$P_0(T)\sim \exp (-\theta_0 T)\sim t^{-\theta_0}$ for large $t$, can not
be simply extended to the $c\neq 0$ case.

It is nevertheless useful to calculate $\theta_d$ by approximating 
the process by a Gaussian process having the same two-point 
correlator for two reasons. First, comparision between
$\theta_d$ obtained numerically for the actual process and that obtained 
for the corresponding Gaussian process will tell us how important are the
non Gaussian effects. Secondly, there have been some recent 
developments\cite{majsire}
in approximate analytical calculations of the exponent for Gaussian 
stationary processes which one can use in the present context.
Therefore, in the following, our strategy would be to
to estimate the exponent $\theta_d(G)$ (that
characterizes the exponential decay of the distribution of intervals
between successive zero crossings, $P(T)\sim e^{-\theta_d(G) T}$ for large 
$T$)
for the Gaussian process with the correlator $f(T)$ as in Eq. (\ref{auto})
and then compare it with the $\theta_d$ for the actual process.

We first present the numerical results for $\theta_d(G)$ for the
Gaussian process with the correlator as in Eq. 
(\ref{auto}). This is done by constructing a time series having the
same corelation function. It is most conveninent to work in
the frequency domain rather than time domain. Details of this simulation
procedure can be found in Ref.\cite{krug}. The results of $\theta_d(G)$ for
different values of $c$ are presented in the third column of Table (1).
By comparing column (2) and (3), it is evident that
the non-Gaussian effects are indeed quite small, and Gaussian approximation
seems to be quite good.

However, exact analytical calculation of $\theta_d(G)$ even for a Gaussian 
stationary process
with a general corelator $f(T)$ is difficult and remains an unsolved problem 
for many years\cite{slepian, reviews}.
Exact results are known only in a few special cases\cite{slepian, reviews}. 
One such 
case is when $f(T)=e^{-\lambda_0 |T|}$ for all $T$. In this case, the 
Gaussian
process is a Markov process and one can show exactly that $P(T)\sim 
e^{-\theta T}$ for large $T$ where $\theta=\lambda_0$. By looking at $f(T)$
in Eq. (\ref{auto}), we see that for $c=1$, $f(T)=e^{-|T|/4}$ for all $T$.
Therefore, $\theta_d(G)=1/4$ for $c=1$. For $c$ close $1$, say 
$c=1+\epsilon$,
one can use a perturbation theory that has recently been developed to
calculate $\theta_d$ for processes that are close to a Markov
process\cite{majsire,Klaus}. According to this theory, if $f(T)= \exp 
(-\lambda |T|) +\epsilon f_1(T)$,
where $\epsilon$ is small, then the exponent $\theta_d$, to order $\epsilon$,
can be most easily expressed as,
\begin{equation}
\label{perturb}
\theta_d(G) =\lambda\bigl[ 1-\epsilon {{2\lambda}\over {\pi}}\int_0^{\infty}
f_1(T)[1-\exp (-2\lambda T)]^{-3/2}dT\bigr].
\end{equation}
In our case, expanding $f(T)$ in Eq. (\ref{auto}) around $c=1$, we get
$\lambda=1/4$ and $f_1(T)=-{1\over {2}}\sinh (T/2)\exp (-3T/4)$. Using this
in Eq. (\ref{perturb}) and performing the integration, we get, to order
$\epsilon$,
\begin{equation}
\label{value}
\theta_d(G)={1\over {4}} +\epsilon {3\over {32}}.
\end{equation}

This perturbation theory may not give good estimates for $\theta_d(G)$ when
$c$ is far away from $1$. However,
one can use a variational estimate for $\theta_d(G)$ for general $c$. This
variational method was recently developed in Ref.\cite{majsire} by 
mapping the
zero crossing problem to that of the evaluation of ground state energy
of a corresponding quantum problem. This method was used\cite{majsire} to 
approximately calculate the ``static" persistence exponent for the Ising
model both in $1$ and $2$ dimensions. The results were in good
agreement\cite{majsire} with the exact result in $1$-d\cite{derrida2} and 
numerical 
simulations\cite{corn} as well as direct experiment\cite{yurke} in $2$-d. 
This method works for class-1 
Gaussian stationary processes, i.e., when $f(T)=1-a|T|+\ldots$ for small
$T$. Since in our present case, $f(T)$ in Eq. (\ref{auto}) is class-1 for 
any nonzero $c$, one can estimate
$\theta_d(G)$ by using this variational method. This method gives two 
estimates $\theta_{max}$ and $\theta_{Var}$ for the exponent $\theta_d(G)$.
While $\theta_{max}$ is a strict rigorous upper bound for $\theta_d(G)$,
$\theta_{Var}$ gives the best variational estmate. The salient 
features of the variational method and the expressions for 
$\theta_{max}$ and $\theta_{Var}$ are given in the Appendix.

For the process being considered with the correlator as in Eq. (\ref{auto}),
estimates $\theta_{Var}$ and $\theta_{max}$ are presented respectively in 
column (4) and (5) of Table (1) for different values of the parameter $c$. 
Comparing these with columns (2) and (3),
it is clear that the variational approximation gets progressively worse
as $c$ increases. A visual summary of these different measures of the
exponent is given in Fig. (2). 

To summarize, we find that the exponent $\theta_d$ depends continuously
on the ratio $c$ of the diffusion constants. For any arbitrary nonzero
$c$, $\theta_d > \theta_0$, where $\theta_0$ is the corresponding static
persistence exponent. Non-Gaussian effects are found to be quite small.

We now turn to the case when the fluctuating field is the spin field
of the Ising model or in general, the ``colour" field of the $q$-state 
Potts model undergoing $T=0$ coarsening dynamics starting from a random 
initial configuration. The tracer particle, once again, moves diffusively
through the medium with a diffusion constant $D_p$ and whenever the field
that the particle sees at its own location changes sign, the particle dies.
As before, one is interested in calculating the survival probability $P(t)$
of the tracer particle.

In the $q$-state Potts model at $T=0$, the domain walls perform random
walk and whenever two domain walls meet, they either annihilate each
other with probability $1/(q-1)$ or coagulate to form a single wall
with probability $(q-2)/(q-1)$\cite{derrida2, majhuse}. The tracer particle 
dies whenever it crosses path with any domain wall. One expects that the 
survival  probability of the tracer particle will decay as $P(t)\sim 
t^{-\theta_p}$ for large $t$. The exponent $\theta_p(q,c)$ is expected, as
in the diffusion case, to depend continuously on $q$ and the ratio  
$c=D_p/D_f$ where $D_p$ is the diffusion constant of the 
test particle and $D_f$ is that of the domain walls.

This problem has been studied in some detail 
before\cite{krapiv1,monthus}. Let us just summarize here the main
results that are already known. In the limit $q\to 
\infty$, $\theta_p(\infty, c)$ can be computed
exactly by noting that only two domain walls on either side of the tracer
particle actually matter for the calculation of $P(t)$\cite{Fisher}. One 
finds exactly, $\theta_p(\infty, c)={\pi}/\{2{\cos}^{-1}[c/(1+c)
]\}$\cite{Fisher,krapiv1}. Note that for
$c=0$, this reduces to the static persistence exponent, 
$\theta_p(\infty,0)=1$\cite{derrida2}.
In the Ising limit, $q=2$, however, there is no exact result for general
$c$. Only exact result is available for $c=0$, 
$\theta_p(2,0)=3/8$\cite{derrida2}. For general $c$ and $q=2$,
a mean field Smoluchowski type approached was developed\cite{krapiv1}, whose
predictions, $\theta_p(2,c)=\sqrt {(1+c)/8}$ were in good agreement with
numerical simulations\cite{krapiv1}. However, this Smoluchowski approach, 
when extended to large $q$ limit, differed substantially\cite{krapiv1} 
from the exact $q\to \infty$ result. Finally a perturbation theory has
been developed recently by Monthus\cite{monthus}, and the exponent
$\theta_p(q,c)$ has been determined at first order perturbation in
$(q-1)$ for arbitrary $c$ and at first order in $c$ for arbitrary $q$.

We will not study this exponent $\theta_p(q,c)$ in its generality
any further in this section. However, in
the next section, we will study in some detail the special case
$c=1$, as it turns out to be a particularly interesting case from the 
point of view of the ``pattern" persistence problem of the Potts model.

\section{Persistence of a Specific Pattern in the $1$-d Potts Model}
\label{potts}

In section-2, we showed that the survival probability of an ``adaptive" 
test particle is related to the peristence of a specific pattern
namely an original domain in the $T=0$ dynamics of the Potts model.
In this section, we show that the survival probability of a 
``noisy" or ``diffusive"  test particle (studied in section-3) is
also related to the persistence of yet another pattern in the 1-d
Potts model namely the survival upto time $t$ of two adjacent original
domains present in the initial configuration.

Let us consider the zero temperature coarsening dynamics
of the $q$-state Potts model starting from a random initial configuration.
In an infinitesimal time interval $dt$, each spin changes its colour to
that of one of its neighbours selected at random. 
This dynamics can be equivalently formulated in terms of the motions of 
domain walls. The domain walls perform independent random walk and whenever
two walls meet, they either annihilate with probability $1/(q-1)$ or
aggregate to become a single wall with probability 
$(q-2)/(q-1)$\cite{derrida3,majhuse}. The ``static" persistence then is
the probability that a fixed point in space is not traversed by any domain
wall. However a somewhat more natural quantity, in this domain wall 
representation,
is the probability $P_1(t)$ that a given domain wall remains {\em uncollided}
upto time $t$. A little thought shows that this is precisely the survival
probability of two adjacent domains present in the initial configuration.
We show below that $P_1(t)\sim t^{-\theta_1}$ for large
$t$ where $\theta_1(q)$ is a $q$-dependent exponent that is not 
obviously related to any other known exponent via scaling relations. 

In section-2, we considered the exponent $\theta_p(q,c)$ 
characterizing the decay of the survival probability of a diffusing
test particle in the background of the diffusing domain walls of the
Potts model. The parameter $c$ is the ratio of the diffusion constant of the
test particle to that of the domain walls. 
Let us consider the case $c=1$. In this 
case, the test particle can not be distinguished from the other diffusing
domain walls. Since the test particle dies whenever any other domain
wall touches it, it is clear that the survival probability of the test
particle for $c=1$ is precisely the fraction of {\em uncollided} domain walls
in the Potts model and hence $\theta_1(q)= \theta_p(q, c=1)$.

Clearly $\theta_1$ can be exactly determined in the two limits, $q=2$ and
$q\to \infty$. For $q=2$, since there is only annihilation upon contact 
between domain walls, the fraction of uncollided walls is the same as the
density of domain walls that decays as $\sim t^{-1/2}$ for large $t$, and
hence $\theta_1(2)=1/2$. In the $q\to \infty$ limit, by putting $c=1$ in
the exact formula, $\theta_p(\infty, c)={\pi}/\{2{\cos}^{-1}[c/(1+c)
]\}$\cite{Fisher,krapiv1}, one gets $\theta_1=3/2$. For intermediate 
values of $q$, we present numerical results in the column (2) of Table (2). 
It is clear that the exponent $\theta_1(q)$ increases monotonically with $q$.

The exponent $\theta_1$ can be quite easily computed within mean field
theory. Let $N(t)$ and $N_1(t)$ denote respectively the total density of 
domain walls and density of ``uncollided" walls at time $t$. Let $Q_1(t)$
denote the density of domains of size $1$ (here $1$ is the lattice spacing
and hence the smallest interval size). Then $N(t)$ and $Q_1(t)$ are related
via the exact relation,
\begin{equation}
\label{density}
{dN\over {dt}}= -{q\over {q-1}}Q_1.
\end{equation}
However there is no such simple exact relationship between $N_1(t)$ and 
$Q_1(t)$. However, if one neglects correlations, it is easy to write such
a relationship within mean field theory,
\begin{equation}
\label{mft}
{dN_1\over {dt}}=-2({N_1\over {N}})^2Q_1-2{N_1\over {N}}
(1-{N_1\over {N}})Q_1
\end{equation}
where the two terms on the right hand side are self explanatory. 
Eliminating $Q_1$ from the two equations above, we get, $N_1\sim 
N^{2(q-1)/q}$. Using the result, $N(t)\sim t^{-1/2}$, we finally
obtain, $\theta_1(q)=(q-1)/q$, within mean field theory. While the
meanfield answer is exact for $q=2$, it gets worse as $q$
increases as evident by comparision with Table (2). Presumably the
mean field value forms a lower bound to the true exponent value, though
we have not been able to prove it.

However, one can obtain rigorous upper bounds to $\theta_1(q)$ as follows.
This can be done by generalizing the arguments used by Derrida\cite{derrida3}
to obtain upper bounds to the ``static" persistence exponent $\theta_0(q)$.
The argument goes as follows. It was noted by Monthus\cite{monthus} that
the problem of a diffusing tracer particle moving amongst the domain walls
of the Potts model can be mapped to a reaction diffusion problem where
particles are generated from a source, diffuse around and aggregate upon
contact. The only difference from the ``static" case\cite{derrida3} was
that the source is now ``moving". In fact the source diffuses with
the same diffusion constant as the tracer particle. It is then possible
to write the survival probability $P_1(t)$ as\cite{monthus},
\begin{equation}
\label{reaction}
P_1(t)=\sum_1^{\infty}P(m,t)q^{1-m}
\end{equation}
where $P(m,t)$ is the probability of having $m$ particles in the 
corresponding reaction diffusion problem. Writing the above equation for 
$q=q_2$ in the following way, 
$a\geq 1$],
\begin{equation}
\label{jensen}
{P_1(q_2)\over {q_2}}= \sum_{m=1}^{\infty}P(m,t)q_2^{-m}=\sum 
P(m,t)[{q_1}^{-m}]^{{\log q_2}/{\log q_1}},
\end{equation}
and then using Jensen's inequality [$\langle x^a\rangle\geq {\langle 
x\rangle}^a$ for $a\geq 1$ and $x$ a positive random variable] as was
used in the ``static" case\cite{derrida3}, we immediately obtain
the following inequality, 
\begin{equation}
\label{inequality}
\theta_1(q_2)\leq\; \theta_1(q_1){{\log {q_2}}\over {{\log {q_1}}}}
\end{equation}
for $q_2\geq q_1$. For example, using the exact result, $\theta_1=1/2$
for $q=2$ and the above inequality we get,
\begin{equation}
\label{q}
\theta_1(q) \leq {\log q}/{2\log 2}.
\end{equation}
For $q=3$, this gives $\theta_1(3)\leq 0.792481...$, which should be 
compared with its numerical value $0.72\pm 0.005$. For higher values of 
$q$, one can have a numerical estimate of a tighter upper bound
of $\theta_1(q)$ by using the numerical value of $\theta_1(q-1)$ in the 
inequality 
(\ref{inequality}). This would be an improvement over the exact
bound (\ref{q}) obtained by comparing with $q=2$. These improved numerical 
bounds $\theta_1(max)$ are presented in the column (3) of Table (2). 

It was pointed out by Monthus\cite{monthus} that carrying out the same
formalism that led to the exact determination of the static persistence 
exponent $\theta_0(q)$\cite{derrida2} is not so straighforward to compute
the exact value of $\theta_p(q,c)$ for general $c$. However, one may hope
that some special simplifications might occur for $c=1$ leading to
the exact computation of the exponent $\theta_1(q)$ though we have not
succeded yet in that direction.

\section{Summary and Conclusion}
\label{summary}

In this paper, we have studied the persistence of some patterns present 
in the initial configuration of a fluctuating field. It was shown that
some of these pattern persistence problems are related to the survival
probability of a mobile particle launched into the field. By suitably
adjusting the rules of the dynamics of the particle one can study
persistence of different patterns in the underlying field. This led
us to study the PF problem in general. Several special cases were
studied in detail and new results were derived.
  
It is clear from ours as well as other studies that there is
a whole hierarchy of exponents associated with the decay of 
persistence of different patterns in the phase ordering systems. It is
not clear at present whether these exponents are independent of each
other or not. While 
these exponents do not depend on the details of the initial configuration 
(as long as it is short ranged), it is not clear whether they can be 
considered ``universal" and if so, in what sense. For example, if the 
diffusion constant of any 
single domain wall of the Potts model changes slightly, then the exponent
$\theta_1$ characterizing the decay of survival of the wall (probability
that it remains uncollided) also changes. Clearly in this respect
the exponent is non-universal. So the important question that remains
to be answered is: what are the criteria one should use to decide whether
an exponent in the phase ordering dynamics is universal or not?

One of the interesting extensions of the present work would be to study
the PF problem when the fluctuating field is the height of an
interface in or approaching the steady state. The ``static" persistence for 
interfaces have
been recently studied in some detail\cite{krug}. Also anomalous diffusive
behaviour of a tracer particle on a solid-on-solid surface was 
noted\cite{jk} and was attributed to the temporary trapping or burial of the 
particle in the bulk of the crystal. Besides, given that sophisticated
techniques using scannning tunneling
microscope already exist for determining temporal step fluctuations on 
crystal surfaces\cite{bartelt}, it is not unreasonable to hope that 
such techniques may be refined in future to measure 
the survival probabilities of the  ``static" as well as the  ``mobile" 
particle in a fluctuating interface.

Finally it has recently been noted\cite{ron} that ``static" persistence
exponent for the diffusion equation may possibly be measured in dense
spin-polarized noble gases ($3$He and $129$Xe) using NMR spectroscopy and
imaging\cite{tseng}. In these systems the polarization acts like a diffusing 
field. With a slight modification these systems may possibly be used to
measure the persistence of some patterns of the diffusive field 
as discussed in the present paper. 

\subsection*{Acknowledgements}
SNM is grateful to Deepak Dhar for very useful discussions, important 
suggestions and a critical reading of the manuscript.
SJC would like to thank Alan Bray and Tim Newman for discussions,
and acknowledges support from EPSERC. 

\section*{Appendix: The Variational Method}

It was shown in \cite{majsire} that the exponent $\theta_d(G)$
is exactly the ground state energy difference, $\theta_d(G)=E_1-E_0$
between two quantum problems, one with a hard wall at the origin and
the other without the wall. The energy $E_0$ (without wall) can be 
exactly determined,
\begin{equation}
\label{ground}
E_0= {1\over {2\pi}}\int_0^{\infty}\log \bigl( {G(\omega)\over 
{{\omega}^2}}\bigr)d\omega
\end{equation}
where $G(\omega)=1/f(\omega)$ and $f(\omega)$ is the Fourier transform
of the correlation function $f(T)$ of the Gaussian stationary process
normalized such that, $f(\omega)\sim {\omega}^{-2}$ for large $\omega$.
The energy $E_1$ (with a wall at the origin)
is estimated variationally as it is hard to obtain exactly. For Class-1 
processes, one can use
a harmonic oscillator with a wall as the trial state with the frequency
$\omega_0$ of the oscillator as the variational parameter. The variational
energy $E_1(\omega_0)$ is given by the expression\cite{majsire},
\begin{equation}
\label{variation}
E_1(\omega_0)=\omega_0\bigl[ {3\over {2}}+{2\over {\pi}}\bigl({G(0)\over 
{{\omega_0}^2}}-1\bigr)+{2\over 
{\pi}}\int_0^{\infty}dx\bigl({G(x\omega_0)\over 
{\omega_0^2}}-x^2-1\bigr)S(x)\bigr],
\end{equation}
where $S(x)=\sum_{n=1}^{\infty}nc_n/(x^2+4n^2)$ with $c_n$'s given by,
\begin{equation}
\label{cn}
c_n={4\over { \pi 2^{2n}(2n+1)!}}\bigl[{(2n)!\over {n!(2n-1)}}\bigr]^2.
\end{equation}
One then minimizes $E_1(\omega_0)$ with respect to $\omega_0$, and uses
this minimizing frequency $\omega_{min}$ to obtain the variational estimate,
$\theta_{max}= E_1(\omega_{min})-E_0(\omega_{min})$ which also is a rigorous
upper bound to the true exponent $\theta_d(G)$ that characterizes the 
Gaussian
process. However, as argued in \cite{majsire}, one can obtain a better
estimate of $\theta_d(G)$ by using, 
$\theta_{Var}=E_1(\omega_{min})-E_0^{(2)}(\omega_{min})$, where 
$E_0^{(2)}$ is given by
\begin{equation}
\label{e02}
E_0^{(2)}=\omega_{min}\bigl[{1\over {2}}+{1\over 
{2\pi}}\int_0^{\infty}dx\bigl( {G(x\omega_{min})\over { 
{\omega^2_{min} (x^2+1)} } }-1\bigr)\bigr].
\end{equation}

\newpage

\Large

{\bf Tables}

\normalsize

\vspace{1cm}

\begin{tabular}{|l|l|l|l|l|}\hline
c  & $\theta_d(MC)$    & $\theta_d(G)$ & $\theta_{Var}$ & $\theta_{max}$ \\ 
\hline 
0.5  & $0.20 \pm 0.01$  & $0.190 \pm 0.005$  & 0.189 & 0.210 \\ \hline
1.0  & $0.26 \pm 0.01$  & $0.25=1/4$       & 1/4   & 1/4  \\ \hline
2.0  & $0.35 \pm 0.01$  & $0.325\pm 0.005$ & 0.319  & 0.350 \\ \hline
3.0  & $0.42 \pm 0.01$  & $0.389\pm 0.005$ & 0.363  & 0.442 \\ \hline
4.0  & $0.48 \pm 0.01$  & $0.439\pm 0.005$ & 0.396  & 0.528 \\ \hline
5.0  & $0.53 \pm 0.01$  & $0.496\pm 0.005$ & 0.422  & 0.611 \\ \hline
6.0  & $0.58 \pm 0.01$  & $0.527\pm 0.005$ & 0.444  & 0.688 \\ \hline
7.0  & $0.62 \pm 0.01$  & $0.579\pm 0.005$ & 0.463  & 0.776 \\ \hline
8.0  & $0.65 \pm 0.01$  & $0.628\pm 0.005$ & 0.479  & 0.839 \\ \hline
9.0  & $0.69 \pm 0.01$  & $0.694\pm 0.005$ & 0.493  & 0.912  \\ \hline
10.0 & $0.73 \pm 0.01$  & $0.723\pm 0.005$ & 0.505  & 0.984   \\ \hline
\end{tabular}

\vspace{0.5cm}

\noindent
{\bf Table I.} Estimates of the
persistence exponents of a diffusing tracer particle through an external 
field evolving via diffusion equation. For different values of the
ratio $c=D_p/D_f$ as shown in column (1), exponents are obtained (2) by 
direct Monte Carlo 
simulation of the process ($\theta_{MC}$), (3) by simulating a Gaussian
stationary process with the correlator $f(T)= [\cosh (T/2)+c \sinh 
(|T|/2)]^{-1/2}$ ($\theta_{GS}$) and (4) by using variational
estimate $\theta_{Var}$ for the above Gaussian stionary process with
correlator $f(T)$ (5) the rigourous upper bound $\theta_{max}$ for the
above Gaussian process. 
Monte Carlo simulations were carried out on a 
periodic lattice of $100000$ sites and the results were averaged over $20$
samples.

\newpage

\normalsize
\vspace{1 cm}

\begin{tabular}{|l|l|l|} \hline

$q$  & $\theta_1$ & $\theta_1(max)$   \\ \hline
2         & 1/2             & -    \\ \hline
3         & $0.72 \pm 0.01$ & 0.792481..  \\ \hline
4         & $0.86 \pm 0.01$ & $0.91 \pm 0.01$  \\ \hline
5         & $0.95 \pm 0.01$ & $1.00 \pm 0.01$  \\ \hline
6         & $1.04 \pm 0.01$ & $1.06 \pm 0.01$  \\ \hline
50        & $1.47 \pm 0.01$ & -  \\ \hline
$\infty$  & 3/2             & -   \\ \hline
\end{tabular}

\vspace{0.5cm}

\noindent
{\bf Table II.} The exponent $\theta_1$ that
characterizes the asymptotic decay $P_1(t)\sim 
t^{-\theta_1}$, probability that a domain wall remains uncollided upto
time $t$ in the zero temperature dynamics of the $q$-state Potts model.
Exact values of $\theta_1$ are quoted for $q=2$ and $q\to \infty$. For
other values of $q$, $\theta_1$ (column (2)) is estimated from Monte Carlo 
simulations
on a periodic lattice of $75000$ sites and results averaged over $20$
different initial conditions. The estimated upper bounds for $\theta_1(q)$
(as explained in the text) are presented in column (3). 

\newpage

\large
{\bf Figure captions}
\normalsize

\vspace{0.5cm}

\noindent
{\bf Figure 1.} 
Log-log plot of Monte Carlo simulations of the ``adaptive" persistence, 
$P_{ad}(t)$ (plus symbols) and ``static" persistence, $P_0(t)$ (cross)
versus time $t$. The simulations were carried out on a periodic lattice
of $100000$ sites and results were averaged over $20$ samples. The
best fit to the straight lines gives the exponent values, 
$\theta_{ad}=0.091\pm 0.002$ and $\theta_0=0.12 \pm 0.001$
 
\vspace{0.5cm}

\noindent
{\bf Figure 2.}
A visual summary of the different measures of the exponent $\theta_d$
as given in Table 1.


\newpage



\newpage

\begin{thebibliography}{99}

\bibitem{derrida1}
B. Derrida, A.J. Bray and C. Godr\`eche, J. Phys. A {\bf 27}, L357 
(1994); D. Stauffer, J. Phys. A {\bf 27}, 5029 (1994); 
B. Derrida, P.M.C. de Oliveira and
D. Stauffer, Physica {\bf 224A}, 604 (1996); E. Ben-Naim, L. Frachebourg 
and P.L. Krapivsky, Phys. Rev. E {\bf 53}, 3078 (1996).

\bibitem{derrida2}
B. Derrida, V. Hakim and V. Pasquier, Phys. Rev. Lett. {\bf 75}, 751 
(1995), and J. Stat. Phys. {\bf 85}, 763 (1996).

\bibitem{derrida3} B. Derrida, J. Phys. A {\bf 28}, 1481 (1995).

\bibitem{diffusion}
S.N. Majumdar, C. Sire, A.J. Bray and S.J. Cornell, Phys. Rev. Lett.
{\bf 77}, 2867 (1996);
B. Derrida, V. Hakim and R. Zeitak, ibid. 2871.

\bibitem{watson} A. Watson, Science, {\bf 274}, 919 (1996).

\bibitem{krapiv1} P.L. Krapivsky, E. Ben-Naim and S. Redner, Phys. Rev. E
{\bf 50}, 2474 (1994).

\bibitem{cardy} J. Cardy, J. Phys. A {\bf 28}, L19 (1995); E. Ben-Naim, 
Phys. Rev. E {\bf 53}, 1566 (1996); M. Howard, J. Phys. A {\bf 29}, 3437 
(1996); S.J. Cornell and A.J. Bray, Phys. Rev. E {\bf 54}, 1153 (1996).

\bibitem{monthus}
C. Monthus, Phys. Rev. E {\bf 54}, 4844 (1996).

\bibitem{krug} J. Krug, H. Kallabis, S.N. Majumdar, S.J. Cornell, A.J. 
Bray and C. Sire, to appear in Phys. Rev. E (1997), {\it cond-mat} 
9704238. 

\bibitem{yurke} B. Yurke, A.N. Pargellis, S.N. Majumdar and C. Sire,
Phys. Rev. E {\bf 56}, R40 (1997).

\bibitem{Tam} W.Y. Tam, R. Zeitak, K.Y. Szeto and J. Stavans, Phys. Rev. 
Lett. {\bf 78}, 1588 (1997).

\bibitem{jk} J. Krug and H.T. Dobbs, Phys. Rev. Lett. {\bf 76}, 4096 (1996);
P.C. Searson, R. Li and K. Sieradzki, Phys. Rev. Lett. {\bf 74}, 1395 (1995).

\bibitem{slepian}
D. Slepian, Bell Syst. Tech. J. {\bf 41}, 463 (1962).

\bibitem{reviews} M. Kac, SIAM Review {\bf 4}, 1 (1962);
I.F. Blake and W.C. Lindsey, IEEE Trans. Information Theory {\bf 19},
295 (1973).

\bibitem{majsire} S.N. Majumdar and C. Sire, Phys. Rev. Lett. {\bf 77},
1420 (1996); C. Sire, S.N. Majumdar and A. Rudinger, unpublished.

\bibitem{Supriya} S. Krishnamurthy and M. Barma, Phys. Rev. Lett. {\bf 76},
423 (1996).

\bibitem{Bak} P. Bak and K. Sneppen, Phys. Rev. Lett. {\bf 71}, 4083 (1993).
 
\bibitem{corn} S.J. Cornell, unpublished.

\bibitem{majdhar} S.N. Majumdar and D. Dhar, unpublished.

\bibitem{krapiv2} P.L. Krapivsky and E. Ben-Naim, {\it cond-mat} 9705155.

\bibitem{Klaus} K.\ Oerding, S.J.\ Cornell, and A.J.\ Bray,
Phys. Rev. E {\bf 56}, R25 (1997).

\bibitem{krapiv3} P.L. Krapivsky and S. Redner, {\it cond-mat} 9603176;
{\it cond-mat} 9502041, 9505062.

\bibitem{majhuse} S.N. Majumdar and D.A. Huse, Phys. Rev. E {\bf 52}, 270 
(1995); C. Sire and S.N. Majumdar, Phys. Rev. Lett. {\bf 74}, 4321 (1995);
Phys. Rev. E {\bf 52}, 244 (1995).

\bibitem{Fisher} M.E. Fisher and M.P. Gelfand, J. Stat. Phys. {\bf 53}, 
175 (1988); D. ben-Avraham, J. Chem. Phys. {\bf 88}, 941 (1988); D. 
Considine and S. Redner, J. Phys. A {\bf 22}, 1621 (1989).

\bibitem{bartelt} N.C. Bartelt, T.L. Einstein and E.D. Williams,
Surf. Sci. {\bf 312}, 411 (1994).

\bibitem{ron} Ronald L. Walsworth, private communication.

\bibitem{tseng} C.H. Tseng, S. Peled, L. Nascimben, E. Oteiza, R.L. 
Walsworth, and F.A. Jolesz, Harvard-Smithsonian preprint no: 4502, to
appear in J. of Mag. Res., series B, 1997.
 
\end{thebibliography}
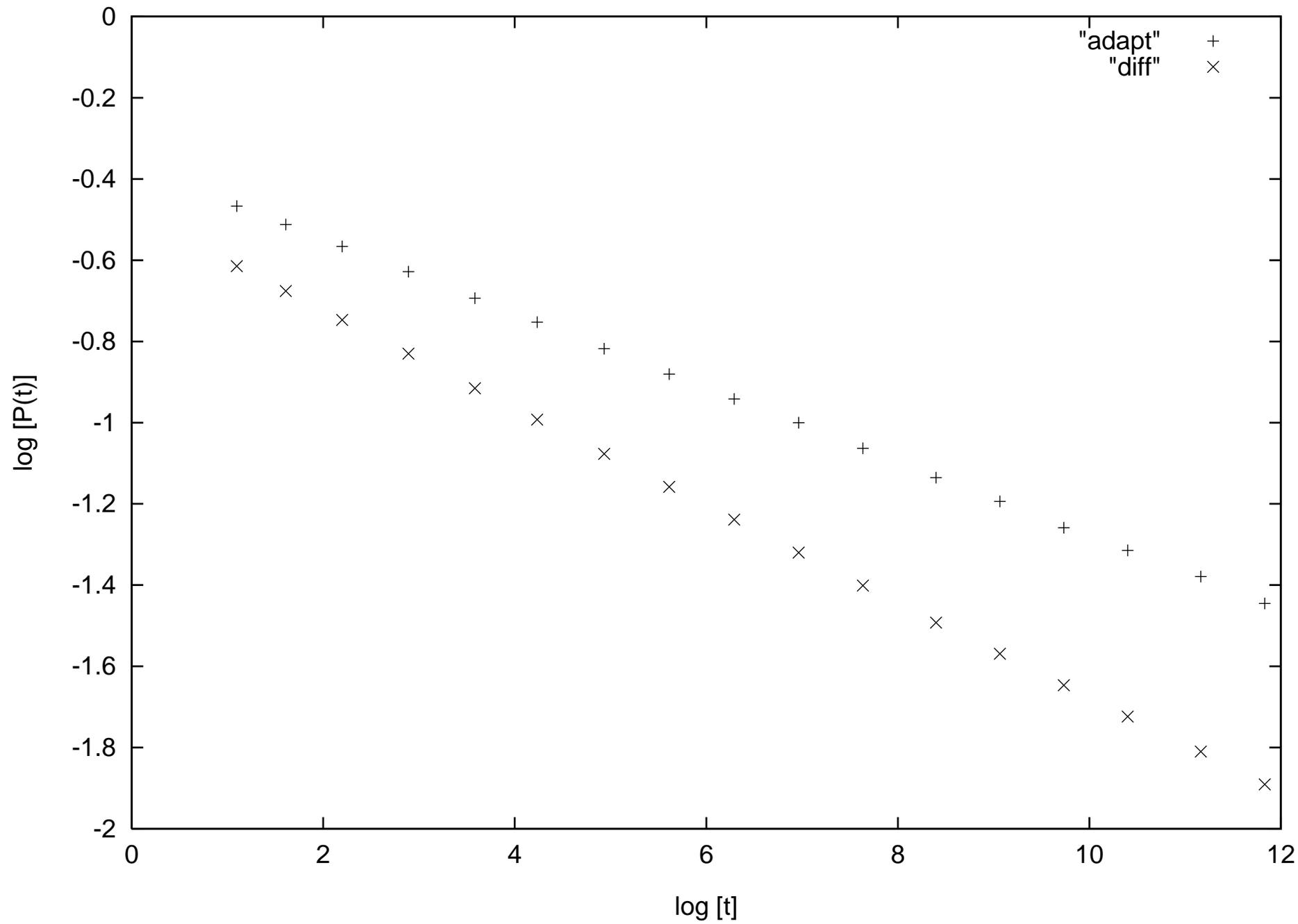
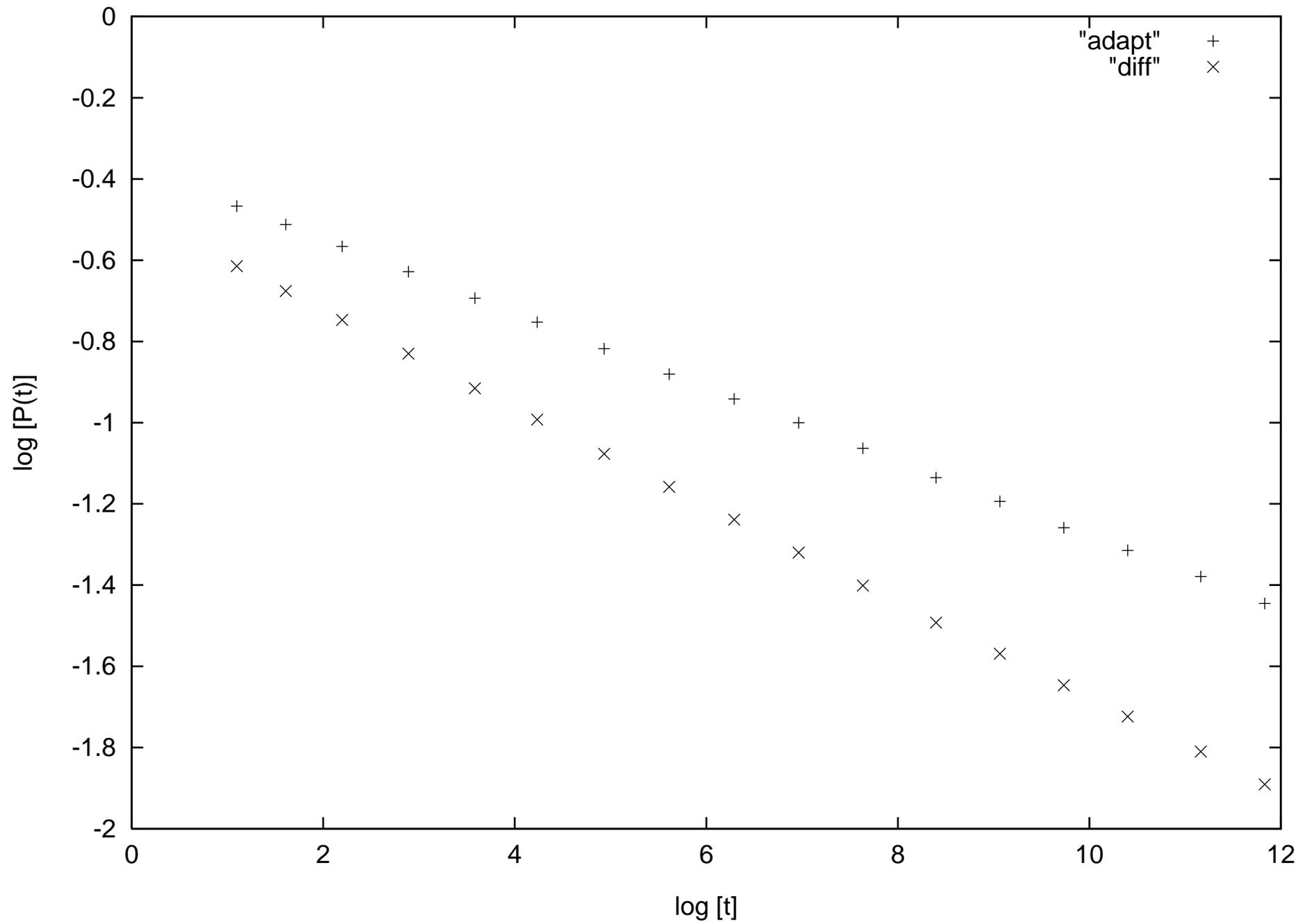
\end{document}